\documentclass[final]{elsart}
\usepackage{graphicx}
\usepackage{amssymb}
\usepackage{amsmath,amsfonts}

\begin{document}
\begin{frontmatter}
\title{On acoustic instability phenomena \\in the vicinity of a lined wall \\exposed to a grazing flow}
\author{Y. Aur\'egan\corauthref{cor}},
\corauth[cor]{Corresponding author.}
\ead{yves.auregan@univ-lemans.fr}
\author{M. Leroux\thanksref{now}}
\thanks[now]{Present address: LEA, Universit\'e de Poitiers, ENSMA, CNRS, Poitiers, France}
\address{%
LAUM, CNRS, Universit\'e du Maine, \\ Av. O. Messiaen, 72085 LE MANS, France}
\begin{abstract}
An experimental investigation of the acoustical behaviour of a liner in a rectangular channel with grazing flow has been conducted. The liner consists of a ceramic structure of parallel square channels: 1mm by 1 mm in cross section, 65 mm in length, and a surface density of 400 channels/inch square. The channels are rigidly terminated, thus constituting a locally reacting structure.
In the absence of flow the liner reacts classically: There is a significant decrease in transmission coefficient around the frequency of minimal impedance. When the wall is exposed to a grazing flow this behaviour is changed:  an increase in transmission coefficient appears at this resonance frequency. The transmission coefficient can be even rise above 1 (up to 3 for a Mach number of 0.3). This behaviour is caused by the appearance of a hydrodynamic instability above the liner.
Furthermore, the stationary pressure drop induced by this liner is deeply affected by its acoustic behaviour. When a sound wave is added, at the resonance frequency of the liner, the pressure drop can increase by a factor 3 when the Mach number is 0.3. This effect is attributed to a modification of the turbulent boundary layer induced by the acoustic wave.
\end{abstract}
\end{frontmatter}
%
\section{Introduction}
\label{Intro}

Acoustically treated ducts are widely used in ducts with flow to reduce noise emission. The calculation of the acoustical propagation in such devices is however difficult because of the complexity of the sound/flow interactions. The coupling between acoustics and flow vorticity can be especially important in the vicinity of a treated wall \cite{auregan01}. Very often in existing models, the flow is simplified and the complexity due to the vortex sound interaction is only taken into account in the Myers condition at the wall \cite{myers80}.

In the case of a perfect fluid with Myers wall condition, Rienstra \cite{rienstra03} has shown that the mode in a flow duct can be split in three different types: An infinite set of acoustical waves, two surface waves (with and without flow), two hydrodynamic surface waves (only with flow). Those surface waves exist only for specific values of the wall impedance. When the hydrodynamic surface waves exist, one of these modes is instable. The recent theoretical interest in this liner instability problem is reflected by the papers of Rienstra \cite{rienstra07} and of Brambley et Peake \cite{brambley06} where a more complete bibliography can be found. 

There has been uncertainty about the physical reality of such instable modes. Experimental demonstration of their existence has however been made by Brandes and Ronneberger \cite{brandes95} in the case of a cylindrical liner. This work was further developed by the G\"{o}ttingen research group (see \cite{Juschke04} for a review of these works). These experimental investigations demonstrate an increase in the acoustic transmission with flow only in the vicinity of the liner resonance. The transmission coefficient can become bigger than one (sound amplification). This effect is associated with a variation of the static pressure drop leading to the possibility of a flow control by the acoustical waves.

In this paper, another experimental evidence of an acoustical instability over an impedance wall in a flow duct is given when the geometry of the duct is two-dimensional. In the first part, the experimental setup is described in detail. The experimental results are given and discussed in the second part. These results clearly demonstrate the existence of an instability leading to a paradoxical increase in the transmission coefficient near the resonance of the liner. The effects of this instability on the static pressure drop of the liner, induced by the wall friction, are also investigated.
\section{The experimental setup}

\begin{figure}[h]
\center
\includegraphics[width=125mm,keepaspectratio=true]{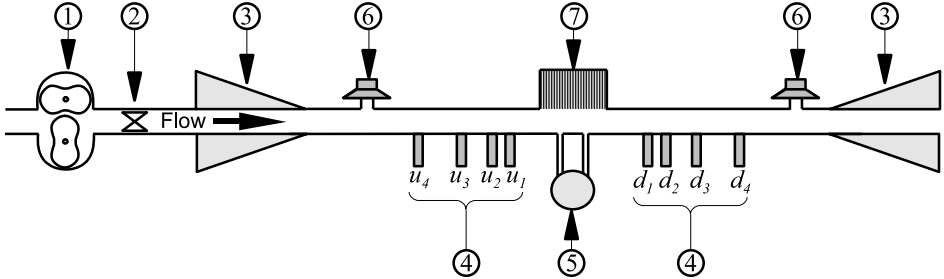}
\caption{Schematic description of the experimental setup.
\textbf{1}:~Compressor, \textbf{2}:~Flowmeter, \textbf{3}:~Anechoic terminations,
\textbf{4}:~Microphones , \textbf{5}:~Static pressure measurement, 
\textbf{6}:~Acoustical source, \textbf{7}:~Lined wall.}
\label{fig:F1}
\end{figure}

The experimental setup is described in Figure \ref{fig:F1}. 
Acoustic waves produced by the two loudspeakers \textbf{(6)} propagate over a liner \textbf{(7)} attached to the wall of a rectangular duct. A grazing flow is superimposed to these waves. The two aims of this setup are: 
\begin{itemize}
\item to measure the acoustic attenuation of the lined wall with a grazing flow with the eight microphones \textbf{(4)}; 
\item to measure the effect of acoustic waves on the static pressure drop induced by the lined wall with the static pressure sensor \textbf{(5)}.
\end{itemize}

\subsection{Acoustical measurements}
\begin{figure}[h]
\center
\includegraphics[width=75mm,keepaspectratio=true]{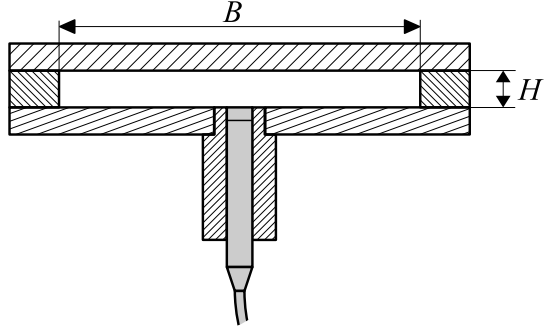}
\caption{Description of microphone setting.}
\label{fig:Sec}
\end{figure}

The air flows from a compressor in an anechoic termination located upstream of the acoustical measurement zone. This termination is comprised of a tube perforated in a non uniform way in order to avoid the reflections of the waves coming from the measurement zone. The tube is covered with a resistive screen and enclosed in a volume to avoid any leakage.
The acoustic measurement zone is rectangular ($H=$ 15 mm by $B=$ 100 mm). This zone is made, see Fig \ref{fig:Sec}, of iron plates (10 mm by 140 mm) screwed to smaller plates (15 mm by 20 mm). These channels have a smooth inner wall with a roughness of less than 0.1 $\mu$m. The duct is 2 m long upstream of the 4 microphones to allow a fully developed turbulent flow. The temperature is measured by two sensors located on both side of the lined sample.

The acoustic source is made with one loudspeaker and one compression chamber allowing an acoustical level of 140 dB in the measurement zone over a wide frequency range (70-3000 Hz). Downstream of the measurement zone, another acoustic source and another anechoic termination are found. In the frequency range 70-3000 Hz, only 2 acoustic modes can propagate: the plane wave and the first order mode along dimension $B$. The microphones being located just at the centre of dimension $B$, this second mode is not measured and the microphones only capture the plane waves (see Fig \ref{fig:Sec}).

The acoustical pressure measurement is performed with two series of four microphones $u_i$ et $d_i$ (microphones \textsl{B\&K 4938} (1/4"), preamplifier \textsl{B\&K 2670} with \textsl{Nexus}). Those microphones are located at the positions, see Fig. \ref{fig:F1}: $x_{u_1} - x_{u_2} = x_{d_1} - x_{d_2} = 63.5$~mm, $x_{u_1} - x_{u_3} = x_{d_1} - x_{d_3} = 211.5$~mm, $x_{u_1} - x_{u_4} = x_{d_1} - x_{d_4} = 700$~mm. 
The use of 2$\times$4 microphones allows an over-determination of the transmitted and reflected waves on both sides of the sample and avoids the problems in the precision of measurement when the acoustic wavelength is close to half the distance between two microphones. All the microphones are calibrated in a separated device relatively to the microphone $u_1$ which is used as the reference. The microphones are mounted flush to the wall without their protective grids to avoid any discontinuity (see Fig \ref{fig:Sec}).

The microphones signals are recorded by an acquisition system \textsl{HP 3565}. This system is used in sweep sinus mode and an average is made over 1000 cycles of the signal. The system drives the acoustical sources on either side of the two sets of microphones. 

\subsection{Method for the measurement of the acoustic scattering matrix}

This experimental apparatus was designed to measure the scattering matrix in pressure for the plane waves of the lined wall with mean flow. 
The scattering matrix for the plane waves $\mathsf{S}$ relates the scattered pressure amplitudes $p_2^+$ and $p_1^-$ to the incident pressure amplitudes $p_1^+$ and $p_2^-$ by
\begin{equation}
\left(\begin{array}{c}{p_2^+} \\
{p_1^-}\label{eq:scat_plan}
\end{array}\right)
= \left[\begin{array}{cc}{T^+}&{R^-}\\
{R^+}&{T^-} \end{array}\right]
\left(\begin{array}{c}{p_1^+} \\ {p_2^-}
\end{array}\right) = \mathsf{S}
\left(\begin{array}{c}{p_1^+} \\ {p_2^-}
\end{array}\right)
\end{equation}
where $T^+$ and $T^-$ are the anechoic transmission coefficients, $R^+$ and $R^-$ are the anechoic reflection coefficients, and the subscripts $i=1,2$ indicate the inlet and the outlet of the lined wall respectively and the superscripts $\pm$ indicate the direction of propagation along the $x$ axis. The paper of {\AA}bom \cite{abom91} reviews the ways to measure these matrices.
The method of measurement used in the present study is called "the 2-sources method". Two measurements are made in two different states of the system. These different states are obtained by switching on the upstream source, the downstream source being switched off (measurement $I$), and vice versa (measurement $II$).

When the two measurements are done, the scattering matrix can be obtained with
\begin{equation}
\left[ \begin{array}{cc}
\left({p_1^-}/{p_1^+}\right)^I & \left({p_1^-}/{p_2^-}\right)^{II}\\ \left({p_2^+}/{p_1^+}\right)^I & \left({p_2^+}/{p_2^-}\right)^{II}
\end{array} \right] =\mathsf{S} \left[ \begin{array}{cc}
1 & \left({p_1^+}/{p_2^-}\right)^{II}\\ \left({p_2^-}/{p_1^+}\right)^{I}&1
\end{array} \right]
\label{eq:ei2}
\end{equation}
if the determinant of the right hand side matrix does not vanish (the superscripts $I$ and $II$ indicate that the quantity had be determined during the measurements $I$ and $II$). This condition $(p_2^-/p_1^+)^{I} \neq (p_1^+/p_2^-)^{II}$ is the condition of independence of the two measurements.
The coefficients of the matrix in Eq. (\ref{eq:ei2}) can be found from the transfer functions between the different microphones by a relation of the type:
\begin{equation}
(p_1^-/p_1^+)^I = \frac{H^I_{u_ju_i}e^{-j k^+x_{u_i}} - e^{-j k^+x_{u_j}}}
{e^{j k^-x_{u_i}} - H^I_{u_ju_i} e^{j k^-x_{u_j}}}
\label{eq:ei3}
\end{equation}
where $H^I_{u_ju_i}$ is the transfer function between the microphones $u_j$ and $u_i$ obtained in the measurement $I$, $k^+$ and $k^-$ are the wavenumbers in the duct in the direction of the flow and in the reverse direction and $x_{u_i}$ is the position of the microphone $u_i$ relatively to the inlet of the measured element. All the other matrix elements can be found on the same way (see \cite{auregan03} for details). Thus, the wavenumbers $k^+$ and $k^-$ have to be known to calculate the scattering matrix.

\subsection{Flow measurements}

\begin{figure}[h]
\center
\includegraphics[width=105mm,keepaspectratio=true]{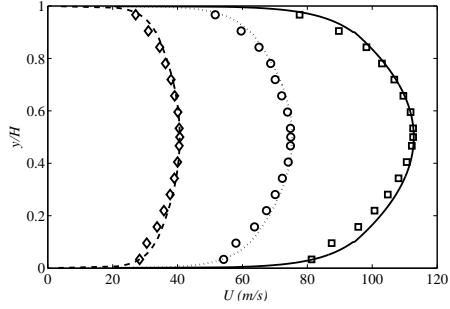}
\caption{Flow profiles for 3 different flow rate. Symbols: Pitot tube measurements, lines: Fully developed profile. The mean Mach numbers are 0.1, 0.2, 0.3}
\label{fig:flow}
\end{figure}

The flow source is a compressor \textsl{Aerzen Delta blower GM10S} that can provide a flow rate up to 0.15 m$^3$.s$^{-1}$. This flow rate is measured with a flow-meter \textsl{ITT Barton 7402} (diameter = 50 mm). The linear range of this device is 0.03 to 0.157 m$^3$.s$^{-1}$. This flow-meter is associated with a measurement of the absolute pressure and of the temperature in order to deduce the mass flow rate.
The flow profile was measured with a pitot tube (diameter 1 mm). This measurement was performed just upstream of the liner. The results for three different flow rate, corresponding to a mean Mach number of 0.1, 0.2, 0.3, are depicted in Figure \ref{fig:flow}. They are compared to the theory for a fully developed turbulent flow in a 2D channel \cite{schlichting79}. The agreement is good enough to consider that the flow is a fully developed turbulent flow.

The static pressure measurement is performed with EFFA GA064A5-20 pressure transducer (range 0--20~ mb). Due to the disturbance induced by this transducer, the acoustic measurements are made when the static pressure transducer is removed. During the static pressure measurement, sound waves are produced by the upstream loudspeaker and the acoustic pressures are measured.

\subsection{The ceramic liner}

\begin{figure}[h]
\center
\includegraphics[width=85mm,keepaspectratio=true]{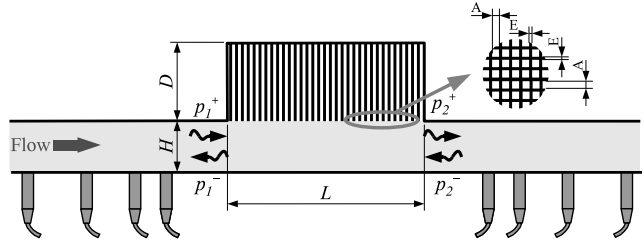}
\caption{Schematic view of the experimental set up with a ceramic liner.}
\label{fig:cera}
\end{figure}

\begin{figure}[h]
\center
\includegraphics[width=60mm,keepaspectratio=true]{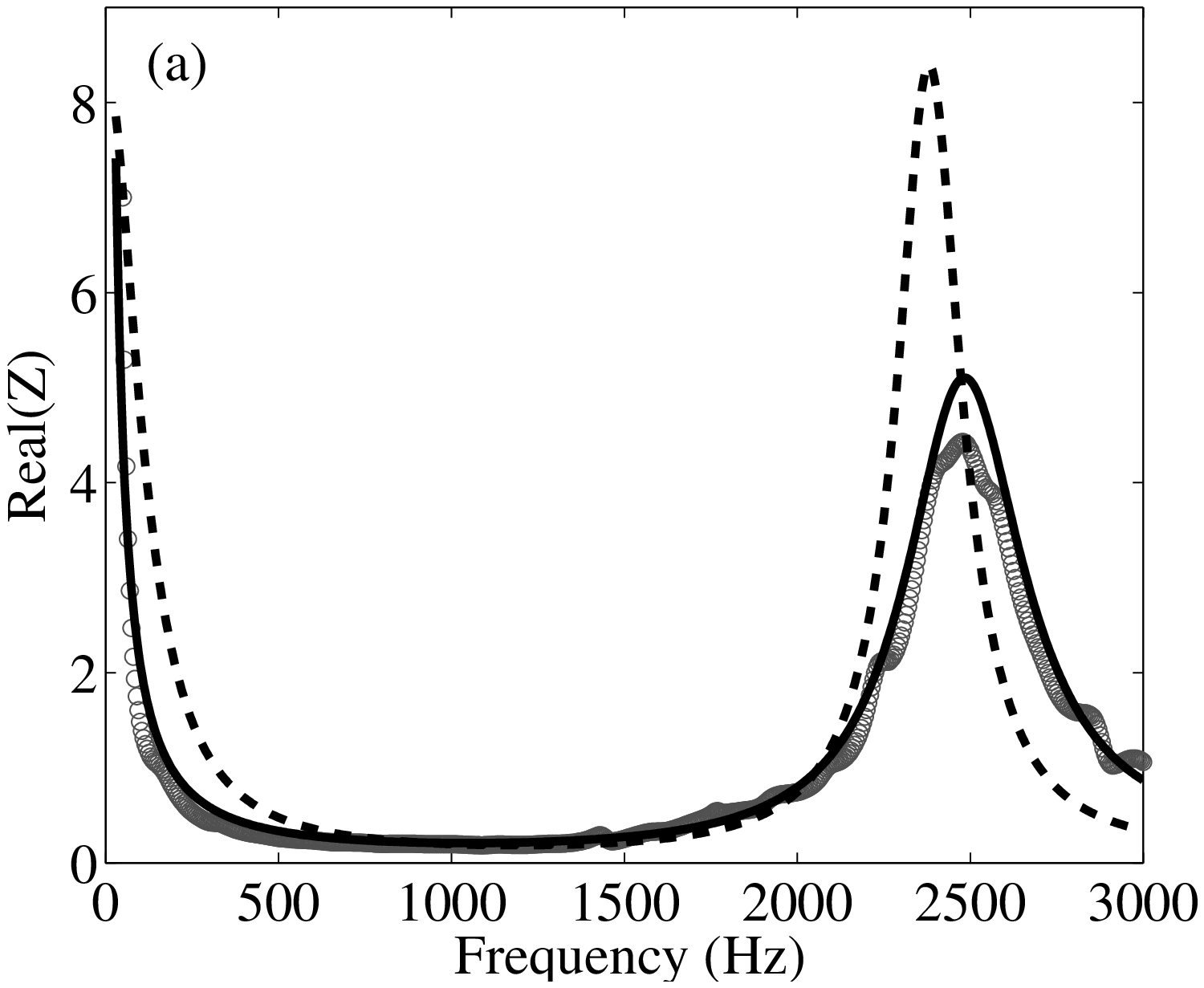}
\includegraphics[width=60mm,keepaspectratio=true]{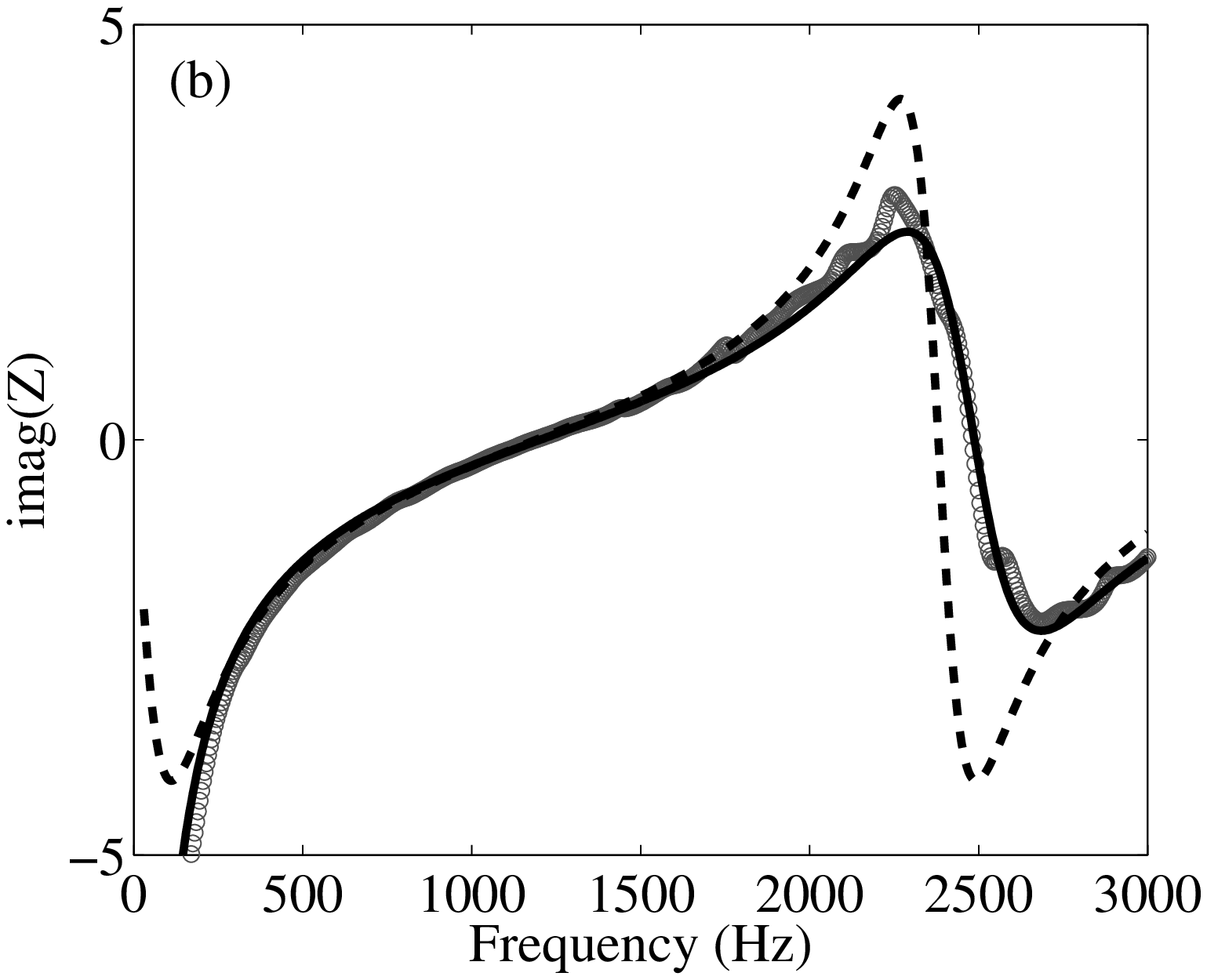}
\caption{Real (a) and imaginary (b) parts of the reduced impedance of the liner measured in an impedance tube (symbol), fitted by Eq. (\ref{eq:imped}) (continuous line) and fitted by Eq. (\ref{eq:imped2})  (dashed line).}
\label{fig:Ztube}
\end{figure}

The tested liner (see Fig. \ref{fig:cera}) is a ceramic structure of parallel and square channels: $A$ = 1 mm by $A$ = 1 mm, and a surface density of 400 channel/inch$^2$ (This liner is typically found in catalytic converters). The channels are rigidly terminated and ensure a locally reacting structure. The depth of the channels is $D$ = 65 mm and the length of the liner is $L$ = 100 mm.
The impedance of this liner was measured in a conventional normal incidence impedance tube (see Fig. \ref{fig:Ztube}). By varying the level during these impedance measurements, the linearity of this ceramic liner was verified. The reduced impedance can be fitted in the frequency range of interest either by an expression derived from the impedance of tube with visco-thermal losses:
\begin{equation}
Z_f = \frac{-j}{\phi} \cot( D \omega /c_0 +(1-j)\alpha \sqrt{\omega}) \label{eq:imped}
\end{equation}
with the porosity $\phi$ = 0.8, $D/c_0$ = 1.85~10$^{-4}$, $\alpha = 2~10^{-3}$ and $\omega = 2 \pi f$ where $f$ is the frequency; 
either by an expression having a simple time domain formulation \cite{rienstra06}:  .
\begin{equation}
Z_f =\frac{-j}{\phi} \cot( D \omega /c_e -j \epsilon /2) \label{eq:imped2}
\end{equation}
with $D/c_e$ = 2.1~10$^{-4}$, $\epsilon = 0.3$.

\section{Results}

\subsection{The acoustic "hump" effect}

\begin{figure}[h]
\center
\includegraphics[width=105mm,keepaspectratio=true]{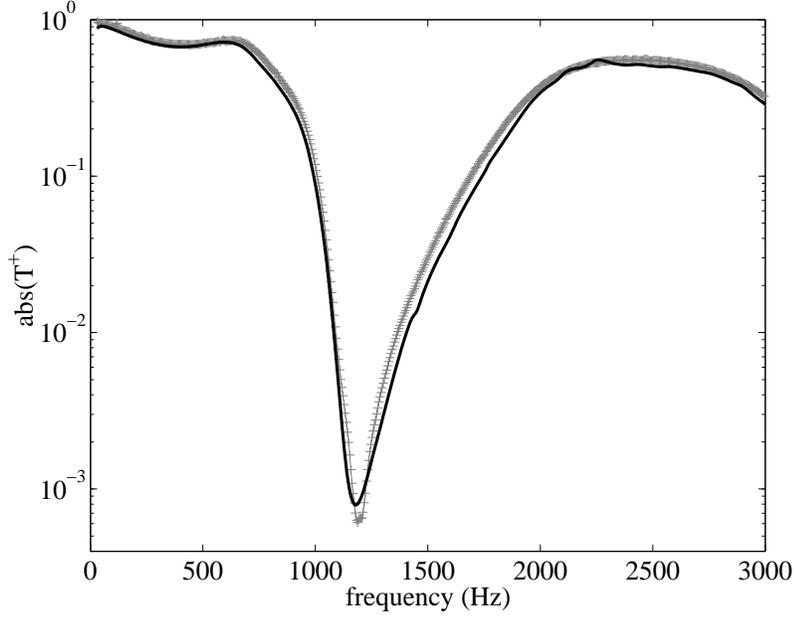}
\caption{Transmission coefficient of the ceramic liner without flow: experimental results (symbols) and computed by a multimodal method from the impedance measured in the impedance tube (line).}
\label{fig:Tse}
\end{figure}

Without flow, the effect of a locally reacting liner is well known. There is a significant drop in the transmission coefficient at the frequency at which the impedance reaches zero (there is a quarter of wave length in the liner). This effect can be seen in Fig. \ref{fig:Tse}. There is a good match between the theoretical prediction, computed with a classical multimodal method with the impedance given by Eq. (\ref{eq:imped}), and the experimental data.

\begin{figure}[h]
\center
\includegraphics[width=105mm,keepaspectratio=true]{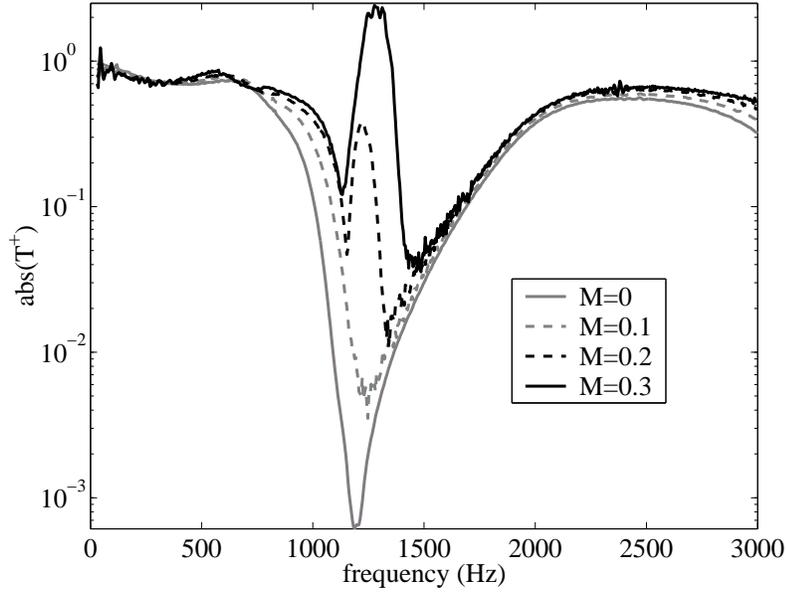}
\caption{Transmission coefficient of the ceramic liner in the flow direction for different Mach numbers $M$.}
\label{fig:Tpe}
\end{figure}

The effect of the flow on this acoustic transmission is now considered.
The transmission coefficient in the flow direction $T^+$ is displayed in Fig. \ref{fig:Tpe} for different Mach numbers $M$. It can be seen that the general shape of the transmission is not greatly affected by flow except near the resonance frequency of the liner where a "hump" appears \cite{auregan11}. Above a certain value of the Mach number, the transmission coefficient can become larger than 1, i.e. the transmitted sound pressure is larger than the incident sound pressure. This effect is not a classical whistling (with a very sharp peak in frequency) but only a sound amplification. This hump effect has been previously seen by Brandes and Ronneberger \cite{brandes95} for a periodic sequence of resonators in a cylindrical duct.

\begin{figure}[h]
\center
\includegraphics[width=115mm,keepaspectratio=true]{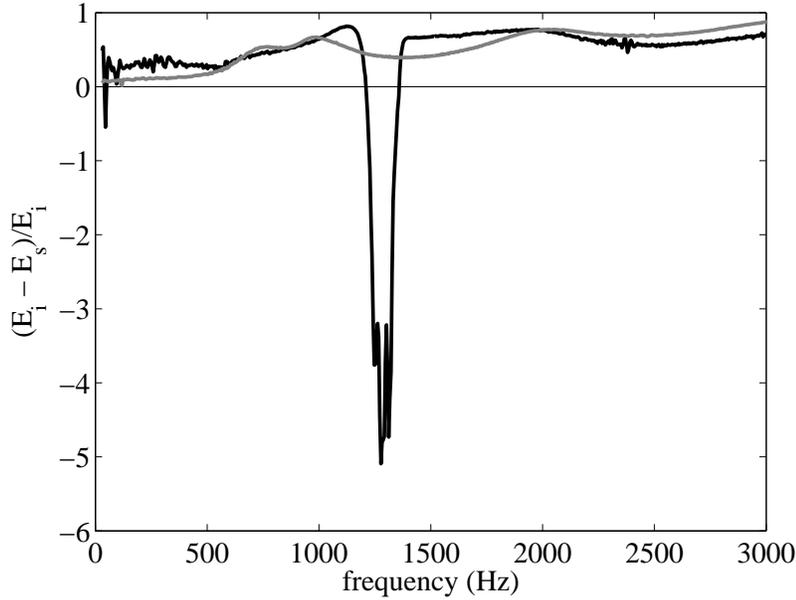}
\caption{Ratio of the dissipated energy (incident energy - scatered energy) on the incident energy. Grey line : $M$ = 0, black line $M$ = 0.3.}
\label{fig:Ene}
\end{figure}

This hump corresponds to an increase in the system's acoustic energy. On Fig. \ref{fig:Ene}, the dissipated energy normalised by the incident energy ($(E_i-E_s)/E_i= 1 - |T^+|^2 - (1-M)^2 |R^+|^2/(1+M)^2$) is plotted \cite{auregan99c}. In the frequency range where the hump is observed, the dissipated energy is less than zero indicating an increase of energy. Thus this hump can be understood as an instability in the system.

\subsection{The hydrodynamic hump effect}

\begin{figure}[h]
\center
\includegraphics[width=105mm,keepaspectratio=true]{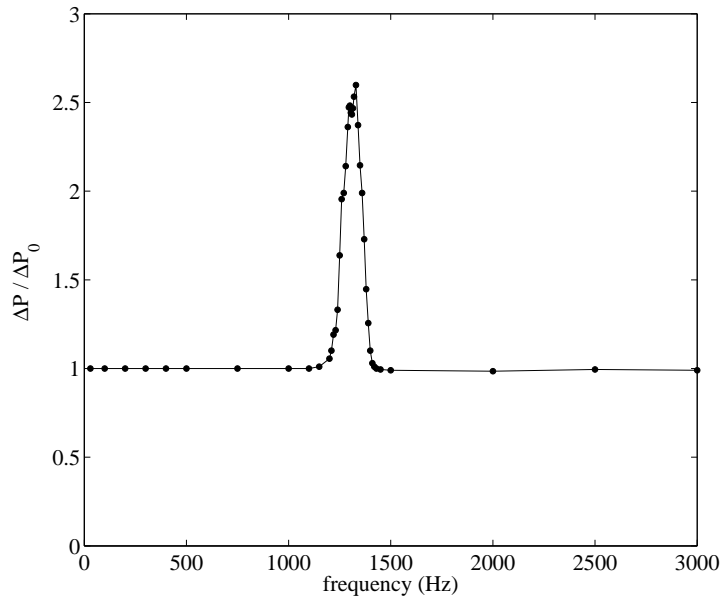}
\caption{Static pressure drop as a function of the frequency of an incident wave on the upstream side with flow (incident level = 130 dB SPL).}
\label{fig:Perte}
\end{figure}

The pressure drop along the liner was measured between two locations lying 6 cm apart. For a given flow, the pressure drop without any acoustic wave was first measured ($\Delta P_0$). Then the pressure drop with an acoustic wave coming from the upstream duct was measured ($\Delta P$). It can be seen from Fig. \ref{fig:Perte} that the acoustic wave induces a very large change in the pressure drop. This surprising nonlinear effect has also been observed by Brandes and Ronneberger \cite{brandes95}.

\begin{figure}[h]
\center
\includegraphics[width=105mm,keepaspectratio=true]{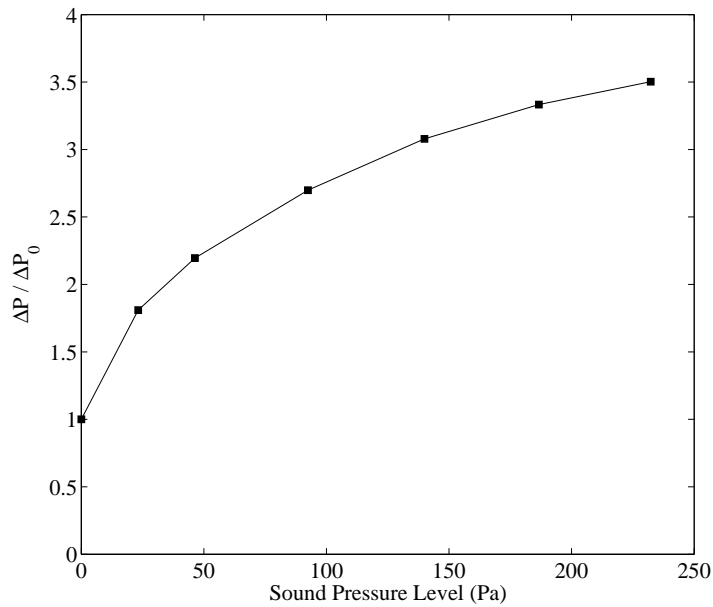}
\caption{Static pressure difference as a function of the sound pressure level of an upstream incident wave with flow.}
\label{fig:Level}
\end{figure}

The value of the rise in pressure difference depends on the value of the Mach number and also on the value of the sound level in the upstream duct (see Fig. \ref{fig:Level}).

\section{Conclusion}

This paper presented experimental evidence of an instability over a lined wall in a two-dimensional duct exposed ot a grazing flow. This hump effect exhibits two related properties: i) the transmission coefficient increase near the resonance of the liner (the coefficient can being larger than one); ii) the mean flow pressure drop is influenced by the level of added acoustic waves. 
From a theoretical point of view, this experimental study could be used as a test of accuracy for computer models of the acoustic propagation in flow. 
Further studies are under way to better understand the role of hydrodynamic unstable modes.
From a practical standpoint, this work will help to better understand the appearance of such a phenomenon and to see if it may happen with usual acoustic liners. 

\bibliographystyle{unsrt}

\end{document}